# Self-assembly of soot nanoparticles on the surface of resistively heated carbon microtubes in near-hexagonal arrays of micropyramids.

Valeriy A. Luchnikov,[1*] Yukie Saito,[2*] Luc Delmotte,[1] Joseph Dentzer,[1] Emmanuel Denys,[1] Vincent Malesys,[1] Ludovic Josien,[1] Laurent Simon,[1] Simon Gree.[1]

[1]Université de Haute-Alsace, CNRS, IS2M, UMR 7361, Mulhouse F-68057, France

[2]Graduate School of Agricultural and Life Sciences, The University of Tokyo, Tokyo, 113-8657, Japan

Corresponding authors: Valeriy A. Luchnikov, email: valeriy.luchnikov@uha.fr and Yukie Saito, email: aysaito@g.ecc.u-tokyo.ac.jp

ABSTRACT Almost regular hexagonal arrays of a few micrometers tall and wide micropyramids consisting of soot nano-particles are formed on the surface of graphitized hollow filaments, which are resistively heated to ∼1800°C−2400°C in an Ar atmosphere containing trace amounts of oxygen (∼300 p.p.m.). At the higher temperatures (T>2300°C, approximately) the soot particles are represented mainly by multi-shell carbon nano-onions. The height and the width of the pyramids is strongly dependent on the temperature of the resistive heating, diminishing from 5-10μm at T≈1800°C to ≈1 μm at 2300-2400°C. Quasi-hexagonal arrays of the micropyramids are organized in the convex "craters" on the surface of the microtubes, which grow with the time of the thermal treatment. The pyramids are pointing always normally to the surface of the craters, except at the boundaries between the craters, where the normal direction is not well defined. The pyramids are soft and can be easy destroyed by touching them, but can be hardened by heating them in the oxygen-free atmosphere. The pyramids are observed only on the exterior surface of the microtubes, but not on their inner surface. This suggests that the thermophoretic force generated by a strong temperature gradient near the external surface of the tubes may be the cause of the micropyramids formation. Electrostatic charging of the soot nanoparticles due to thermionic emission may also be relevant to this phenomenon. The micropyramids can function as field emission point sources, as demonstrated with the use of a micro-nanoprobing station, mounted in a scanning electron microscope.

KEYWORDS: Self-assembly; carbon nano-onions; resistive heating; thermophoretic force; micropyramids; hexagonal array.

Carbon is known for its capacity to form plenty of architectures at the nanoscale as well as the microscale. Apart of the most well known carbon single and multishell tubes, fullerens, and graphene, there have been discovered numerous interesting entities such as carbon nano-horns and nano-cones,[1-3] conical micro-crystals,[4] carbon whiskers,[5,6] graphite micro-pyramides,[7] graphene flowers or vertical graphene.[8-12] Electrical, optical, and morphological peculiarities of these architectures make them perspective for advanced applications such as field electron emission sources,[9,10] supercapacitors,[13] biosensors,[11] ultra-black materials,[14] superhydrophobic[15] and bactericidal[16] coatings. The methods of production of these structures include argon plasma etching of graphite substrates,[7,14] reactive sputtering using methane gas,[9] thermal chemical vapor deposition on carbon fibers,[10] reduction of graphene oxide nanosheets,[11] combustion flame deposition,[8] microwave plasma chemical vapor deposition, followed by bias-assisted reactive ion etching,[16] wood charcoal heat treatment above 2000°C.[6]

Here we report soot micropyramids, which we have found on the surface of resistively heated amorphous carbon microtubes in Ar atmosphere, to which vanishingly small amount of air was admixed. The pyramids are always pointing along the local normal direction to the tube's surface, suggesting that the soot nanoparticles, formed in the oxygen-deficient atmosphere, organize into the pyramids under the action of some force, directed outward the surface. Moreover, the pyramids are assembled in almost hexagonal arrays, resembling the spontaneously formed hexagonal patterns, which appear at certain deformable interfaces under the action of normally applied forces, and which have been extensively studied in the past. Examples include the Rosenweig instability of a ferromagnetic fluid under a magnetic field,[17] the gravity-driven Rayleigh−Taylor instability of suspended liquids,[18] elastic solids,[19] and electrostatically induced structure formations at the polymer−air interface.[20] The patterns arise due to the competition of external fields (gravity, magnetic, or electrostatic), which amplify the interface deformation with surface tension and elastic energy, which have stabilizing effects. Linear stability analyses establish the dispersion relation between the amplitude growth rates of the normal modes of deformation and the wave numbers of the modes. The modes with the highest growth rates dominate and determine the characteristic wavelength of the pattern. Hexagonal patterns arise due to the resonance of the dominant modes, whose wave vectors form an equilateral triangle in the reciprocal space, as shown by nonlinear stability analysis.[18,21]

A natural candidate for the role of such a normally directed force in our system of soot particles on the surface of a resistively heated microtube is the thermophoretic force, $F_{th}$, which results from the difference of the average momentum transferred to the particle by molecules arriving from the forward and the backward directions, with respect to the temperature gradient's direction.[22] More rapid molecules, which arrive from the region of higher temperature, transfer larger momentum on average than the molecules that arrive from the colder regions. This results in a drag force on the particle, which drives it from hotter regions to colder ones. Thermoelectric phenomena, such as electrostatic charging of the particles due to thermionic loss of electrons at high temperatures may be also essential for the process of the micropyramids formation.

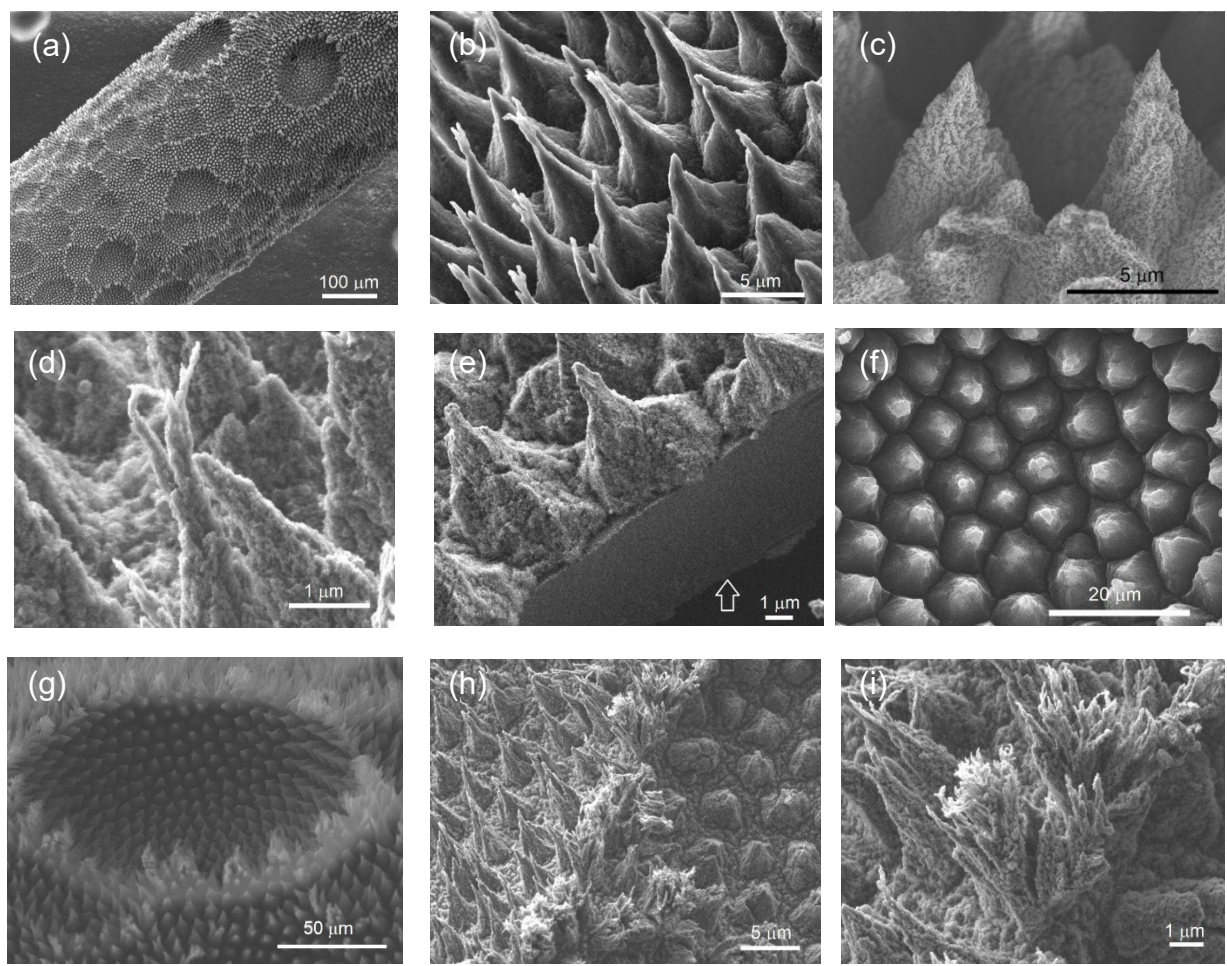

**Figure 1.** The most characteristic morphology features of the carbon microtubes resistively heated in the oxygen-deficient atmosphere. (a) Craters on the surface of a microtube. (b,c) An array of pyramids at different magnification. (d) Almost hexagonal array of the pyramids. (e) The interface between pyramid's bottoms and the unchanged wall of a tube. The white arrow : inner surface of the tube. (f) Pyramids with thin filaments on the top. (g) A magnified view of a crater. Note that the pyramids are pointing normally to the walls of the craters (h) A boundary between two craters. (i) Irregular multi-apex pyramids at the bondary.

**Results and discussion.**

Figure 1 presents the most representative features of the micro-scale morphology observed on the surface of a tube heated to $T = 2000° - 2100°C$ and at an oxygen concentration $\approx$ 300 p. p. m. for 5 hours. The surface of the tube, which was initially smooth, became covered by micro-craters (Figure 1a) whose concave bottoms are roofed by the arrays of the features which we call micro-pyramids, because their shape resembles this geometric form (Figure 1b,c). No similar structure was found on the surface of the control tubes heated without air admixing (see Figure S1 of the Supporting Information, SI). The pyramids have sharp apices with submicrometric curvature radii, and the apices of some pyramids are split. The surface of the pyramids is not smooth, rather, it has porous aspect (see Figure 1c and Figure S2 of SI for a high-resolution image). The pyramids are often prolonged by a thread whose thickness does not exceed 100 nm (Figure 1d). A fragment of an extremity of an intentionally broken tube is shown in Figure 1e. The image reveals the interface between the loose material of the pyramids and the monolithic structure of the tube wall, which was not yet etched by the oxygen-containing atmosphere. The pyramids appear only on the outer surface of the tube but not on their interior walls, shown by the white arrow.

Seen from the top, the pyramid arrays have nearly hexagonal coordination in the centers of the craters (Figure 1f). The pyramids are always oriented normally to the concave craters walls (Figure 1g). At the boundaries between the craters, where the normal direction is not defined, the pyramids have irregular shapes characterized by split apices (Figures 1g,1i).

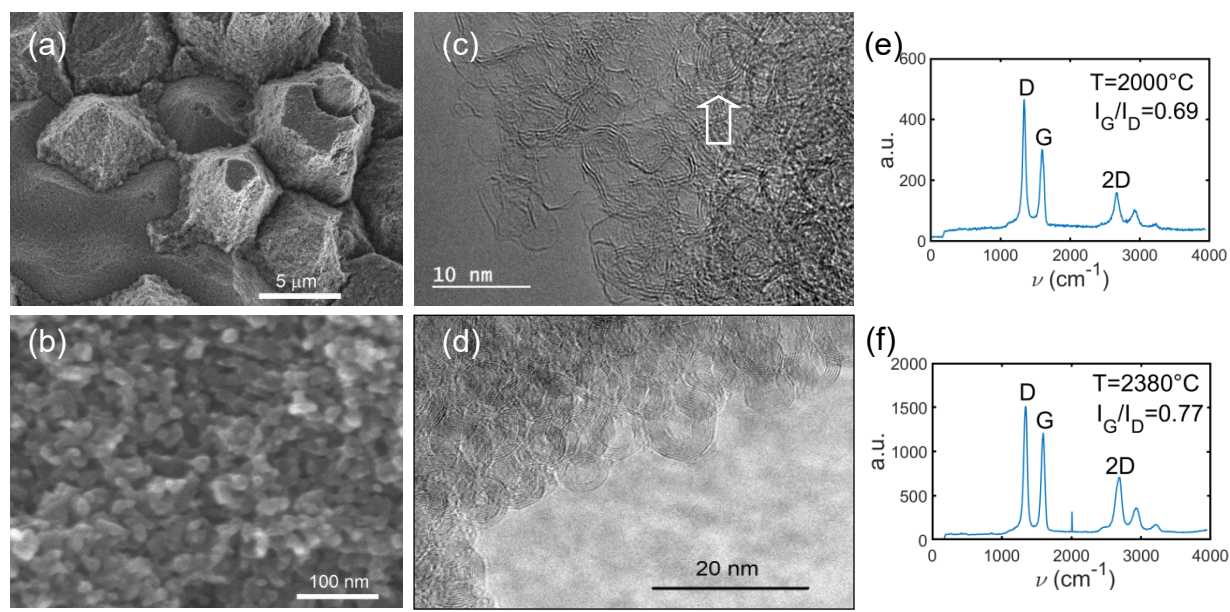

**Figure 2** The inner structure of the pyramids. (a) Pyramids after the application of Scotch tape to the surface of a tube ($T \approx 2000°C$, $C_{O_2} \approx 300$ p.p.m.). Some pyramids are partially destroyed, and some are completely removed. Note that the soot pyramids cover the pyramidal elevations of the unchanged material of the tubes. (b) High-resolution SEM of the interior of a partially destroyed pyramid. (c,d) HRTEM of the material of the miccropyramids, formed at $T \approx 2000°C$ and $T \approx 2380°C$, respectively. A CNO particle is shown on the figure (c) by the white arrow. (e, f) the Raman spectra of the material of the pyramids, formed at $T \approx 2000°C$ and $T \approx 2380°C$, respectively.

To collect the material of the pyramids for Raman spectroscopy characterization, a tube was touched by a narrow strip of Scotch tape. Some of the pyramids were destroyed partially by this operation (Figure 2a), which allowed the examination of their inner structure. High-resolution SEM reveals that the pyramids consist of nanoparticles whose diameters do not exceed 30nm (Figure 2b). The fact that the pyramids can be easily disrupted by touching signifies that the particles are not connected by covalent bonds; rather, they are held together by van der Waals forces. A high-resolution TEM reveals that particles are the closed-shell ones, similar to the particles produced by the carbon arc method.[23] Some particles can be identified as multi-shell carbon nano-onions (CNO) (Figure 2c). The graphitization degree and the fraction of CNOs increase at higher temperatures. For a tube heated to T≈2380°C (Figure 2d) the quasi-totality of the particles, collected on the surface of the tubes, can be classified as CNO. The Raman spectra of the particles (Figure 2e,f) have well-resolved G (graphitic) and D (disordered) bands,[24] meaning that the particles contain a relatively low concentration of defects of the sp2-coordination of carbon atoms. The ratio of the intensities of the G and the D peaks, $I_G/I_D$, grows with temperature, implying the improving the sp2-coordination of the CNO particles.

The mechanism of formation of the soot particles on the surface of the resistively heated graphitized microtubes is not yet clear. The physico-chemical conditions at the surface of the microtubes are different from the conditions of the well-understood processes of soot production as hydrocarbon flames,[25,26] spark discharges[27] and laser ablation.[28] In the flames, the process starts by the gas-phase chemical reactions leading to the formation of polycyclic aromatic hydrocarbons (PAH), followed by the particles nucleation *via* the PAH dimerization, particle growth by the surface reactions, and particles coagulation. It is unlikely that the carbonized microtubes can be the source of PAH precursors, because at temperatures above ≈1000°K carbon films are almost completely dehydrogenated.[29] In the cases of spark discharge and laser ablation, soot nuclei are formed due to evaporation of a small amount of carbon electrode or carbon target material, and condensation of the supersaturated vapor in the neutral gas flow. The soot nuclei coagulate, forming the soot particles, which aggregate eventually. This mechanism is also unlikely be relevant to the formation of the soot particles in our system, because the temperatures of the resistive heating of the microtubes is well below of graphite sublimation point (around 4000°K at the ambient pressure).[30]

To the best of our knowledge, soot formation on the surface of graphitized materials, exposed to the temperatures close to 2000°C and oxygen concentrations of a few hundreds of p.p.m. was not reported to the date, despite a great deal of literature on oxidation of graphite materials, especially for the nuclear applications.[31] Typically, temperature at which graphite oxidation was studied did not exceed 1000°K. At these temperatures, the basal graphite plane is inert to molecular oxygen. The graphite oxidation proceeds from the zig-zag and armchair active sites of the graphite crystallites, attacked by molecular oxygen, with the CO and $CO_2$ desorbtion. Oxidation of multilayer graphene flakes suspended on amorphous carbon grid in air and heated by laser to temperatures close to 2000°K resulted in the layer-by-layer thinning of the flakes, assuming that flakes are eroded in a highly anisotropic manner. Yet, no formation of carbon nanoparticles, such as soot, was reported.[32]

In order to gain a deeper insight into the conditions of the soot particles formation in our system, we have made a control experiment, in which a 25μm-thick pyrolytic graphite film was resistively heated to 2000°C in the Ar atmosphere at oxygen concentration 600 p.p.m. The sample was eroded in course of 20min of the treatment (Figure S3), but no soot and pyramids formation was detected. In particular, the graphite stripe remained shiny, while the carbonized chitosan tubes became violet black upon etching. This means that the mechanism of the soot formation in our system requires a relatively high number of defects of the sp2-coordination of the carbon material.

|  | 1800°C | 1900°C | 2000°C | 2100°C | 2200°C |
|---|---|---|---|---|---|
| $\lambda$ (μm) | $6.2 \pm 0.2$ | $5.8 \pm 0.3$ | $4.3 \pm 0.2$ | $2.8 \pm 0.2$ | $1.0 \pm 0.1$ |

**Table 1.** Average distance between the pyramide centers as the function of resistive heating temperature, for $C_{O_2} \approx 300$ p.p.m.

The size of the pyramids and their number per unit surface depend strongly on temperature of the resistive heating (Table 1). Within the temperature interval 1800°C – 2200°C, in which the micropyramids arrays can be unambiguously identified, the average distance between the pyramids centers decreases by ~6 times as temperature grows. The pyramids also became more « slim », and their apices are more often continued by a thin filament.

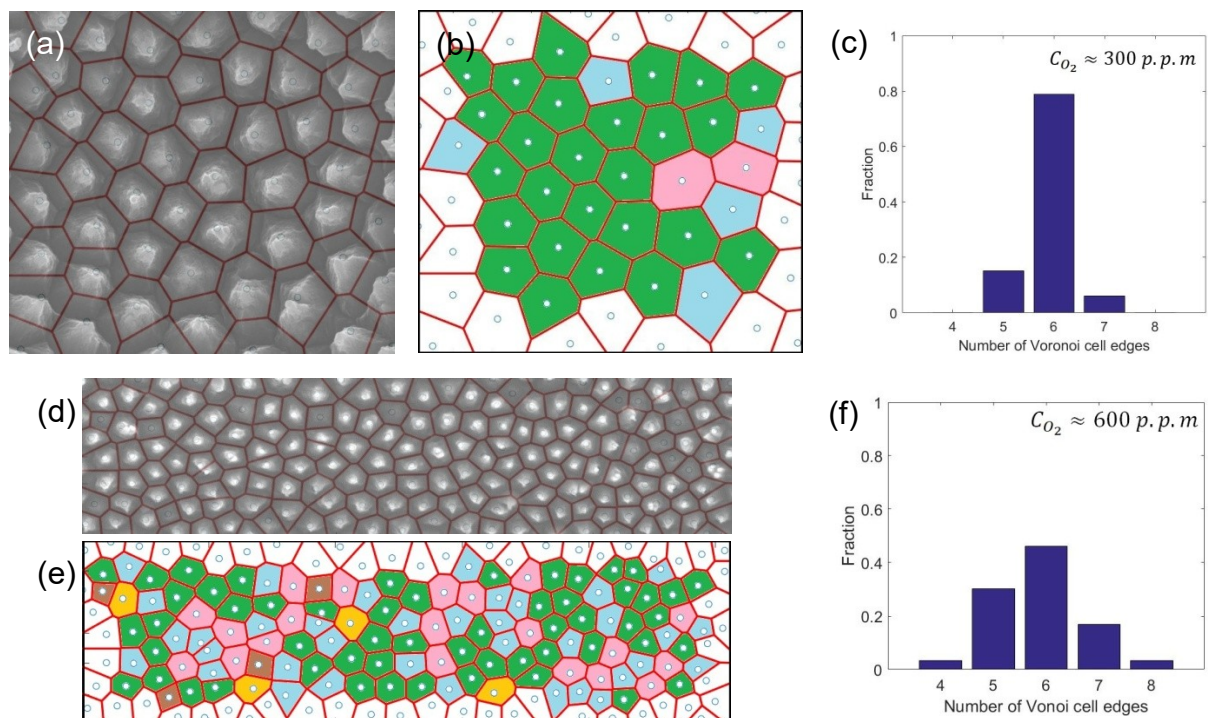

**Figure 3** Voronoi analysis of the pyramids arrangement. Color coding of the Voronoi cells coordination: brown – 4-edged cells; blue – 5-edged cells; green- 6-edged cells; pink – 7-edged cells, and orange – 8-edged cells. (a) Superposition of the SEM image of a crater bottom for a tube heated at $T \approx 2000°C$, and $C_{O_2} \approx 300$ p.p.m during 4 hours, and the Voronoi diagram, calculated by the pyramids vortices. (b) Coloring of the Voronoi diagram cells according to the number of the cells edges. (c) Distribution of the cells by the number of their edges. (d-f) The same as (a-c), respectively, for a tube heated at $T \approx 2000°C$, $C_{O_2} \approx 600$ p.p.m, during 2 hours (see figure S5, SI).

The crater-like features imply that the pyramids do not appear simultaneously on the surface of the tubes. Indeed, we have found that initially small, isolated craters, having an almost perfectly circular shape and containing a few pyramids, appear on the surface of the tubes a few dozens of minutes after the beginning of the resistive heating (Figure S4a,b, SI). During the experiment, the craters became wider and deeper, and the number of pyramids in them increased. New craters appeared and grew. Eventually, all the surfaces became covered by the craters, which touched each other (Figure S4c,d).

The characteristic time of the pyramids formation can be strongly reduced by increasing the oxygen concentration in the vessel atmosphere. Thus, the almost regular arrays of the pyramids were observed on the tubes heated at $T = 2000°C$, $C_{O_2} \approx 600$ p.p.m. after 120 min of resistive heating. A few craters were observed on the surface of the tubes, but the majority of the pyramids were formed outside them (Figure S5, SI). Apparently, the rate of the pyramids formation affects their mutual coordination. This fact follows from the analysis of the Voronoi diagrams of the SEM images of the pyramids arrays produced at at $T = 2000°C$ and at the relative oxygen concentrations $C_{O_2} \approx 300$ p.p.m. and $C_{O_2} \approx 600$ p.p.m (Figure 3). By definition, the Voronoi cell of a vortex is defined by the ensemble of the points which are closer to a given vortex than to any other vortex of the system. The edges of the polygons correspond roughly to the faces of the pyramids, and the vortices of the polygons correspond to the edges of the pyramids. In our analysis, the vortices were defined as the extremities of the micropyramids. At the lower oxygen concentration, the absolute majority (~80%) of the Voronoi cells have 6 edges, while at the higher oxygen concentration, the fraction of the 6-coordinated cells drops to around 50%. Further increase of oxygen concentration to $C_{O_2} \approx 3000$ p.p.m. led to rapid consumption of the microtubes (typically, within 10-20 min) and irregular shapes of the pyramids, as well as poor degree of their hexagonal arrangement.

The hypothetical mechanism of the pyramids growing and the propagation of the pyramids arrays is shown on the Figure 4. We suppose that a pyramid's formation starts from a local defect on the surface of a tube, such as a local microscale elevation (Figure 4a). Soot particles, generated at the bottom of the defect, roll up along its walls in the vertex direction, increasing the defect's height and transforming it into a pyramide. Simultaneously, local cavities are formed at the bottom of the pyramide, because of the particles excavation from the tube. The particles start to roll along the walls of the cavities, forming new pyramids at their borders (Figure 4b). This process is repeated, leading to formation of the pyramid's arrays (Figure 4c). Similar process may start from a defect in form of a micro-cavity on the surface of a tube. The growing of the pyramids seems to be limited by their sputtering from the pyramid's vortices. The formation of the concave "craters" signifies that the pyramids accelerate somehow the transformation of the graphitized material in soot, because the older pyramids have propagated deeper in the walls of the tubes. This is probably because the migration of the soot particles onto the pyramid's walls unmasks the tube's surface and makes it more accessible to the molecules of oxygen present in the gas mixture.

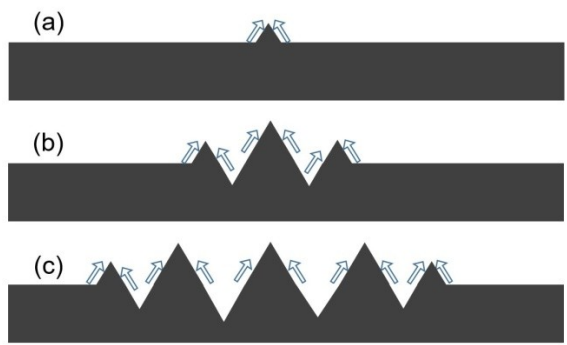

**Figure 4.** Hypothetical mechanism of pyramids growing and pyramids arrays propagation.

The height of the pyramids in the largest craters is much smaller than the craters' depth (see Figure 1g). This means that the soot nanoparticles, which constitute the pyramids, are constantly removed from the pyramids and replaced by new particles, which are produced by the interaction of oxygen with the tube's walls. To clarify the fate of the eliminated particles, we placed a tungsten wire close to a microtube. In 1 hour of resistive heating at $T = 2300°C$, the so-called graphene flowers[8,9] formed on the surface of the wire (Figure S6). Thus, the carbon material did not burn completely; rather, it sputtered into the space around the tubes, where it can be sedimented on a support.

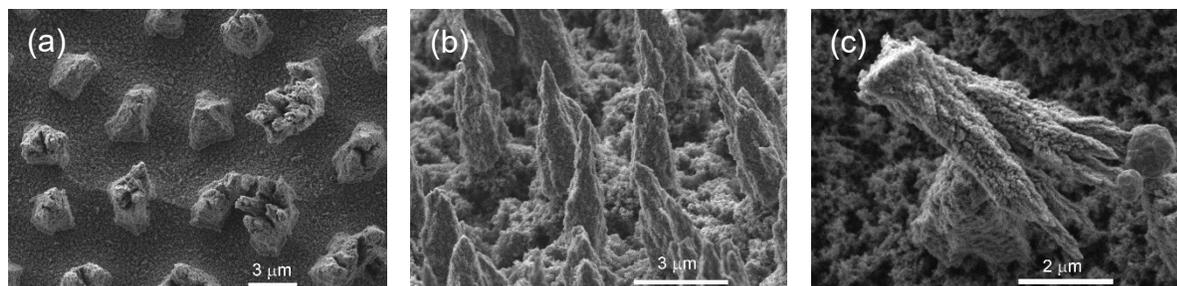

**Figure 5** Normal view (a) and side view (b) of the micropyramids produced at $T \approx 2000°C$, $C_{O_2} \approx 300\,\text{p.p.m.}$ and stabilized by heating at $T \approx 2300°C$ without air admixing during 30min. The sintering of soot particles rigidifies the pyramids, so that they are not smashed or disrupted when touched (cf. Figure 2a); rather, they are wholly unrooted (c).

The pyramids can be stabilized by sintering soot particles in an oxygen-free atmosphere. This operation makes them mechanically stable (Figure 5), so that the pyramids are wholly unrooted rather than smashed, when they are touched. The sintering procedure increases also the height to base aspect ratio of the pyramids, and cleans up the space between them from the soot particles, so that the pyramids are transformed into needle-like structures, well separated from each other. We suppose that the effect of the « cleaning up » of the space between the pyramids is due to the fact that, in the absence of oxygen, soot particles are not produced anymore in course of the erosion of the material of the tube. But they are not sintered immediately, and therefore migrate to the top of the pyramids, increasing their height to width aspect ratio. The particles which are inside the pyramids (and not on their surface) are less mobile, and for them the probability of sintering is higher, therefore these particles constitute the rigid « skeleton » of the pyramids, making them more robust mechanically.

The ensemble of the experimental facts presented above allows us to suppose the existence of a force normally directed to the interface between the soot layer and the material of the tube, which is not yet transformed into soot by its interaction with oxygen. A natural candidate for the role of such a force is thermophoretic force, $F_{th}$, which results from the difference of the average momentum transferred to the particle by molecules arriving from the forward and the backward directions, respecting the temperature gradient's direction. More rapid molecules, which arrive from the region of higher temperature, transfer larger momentum on average than the molecules that arrive from the colder regions. This results in a drag force on the particle, which drives it from hotter regions to colder ones. The fact that the pyramids do not appear inside the tubes, where the temperature is presumably uniform, seems to support the hypothesis that the temperature gradient giving rise to thermophoretic force is the factor that causes the formation of pyramids. The intensity of $F_{th}$ can be estimated from the following considerations. At $T = 2000°C$, the mean free path of argon atoms is $l = k_B(T + 273.15°)/\pi\sqrt{2}pd_{Ar}^2 \approx 435$nm, where $k_B$ is the Boltzmann constant, $p$ is the pressure of the gas, and $d_{Ar} \approx 0.4$nm is the diameter of the argon atoms. From the HRTEM images (Figure 2c,d), one can evaluate the characteristic diameter of the soot particles as $a \approx 10$nm. Since the Knudsen number $l/a \approx 43$ is large, the thermophoretic force can be calculated in the free molecular limit[22] as $F_{th} = -\frac{16\sqrt{\pi}}{15}\frac{a^2\kappa}{\sqrt{2k_BT/m_{Ar}}}\nabla T$ where $\kappa \approx 0.074 W \cdot m^{-1} \cdot K^{-1}$ is argon gas thermal conductivity at the given temperature,[33] $m_{Ar}$ is the mass of argon atoms. The temperature gradient at the filament's surface is usually evaluated with the assumption that there exists a layer of almost immovable gas around the filament (the so-called stagnant layer of Langmuir).[34] Coherent anti-Stokes Raman scattering measurements[35] and computer simulations[36] estimate this gradient to $\sim 10^5 - 10^6 K \cdot m^{-1}$ for tungsten filaments heated to $\sim 2000°K$ at the normal pressure of the nitrogen surrounding the filament. Assuming that, for argon, the temperature gradient is on the same order of magnitude, one can estimate the thermophoretic force acting on a particle of the characteristic size $a \approx 10$nm as $F_{th} \approx 10^{-15} - 10^{-14}$N. It is several orders of magnitude smaller than the typical adhesion force of soot particles measured by atomic force microscopy to be in the range of a few nN.[37] Therefore, it is hardly possible that the thermophoretic force alone can detach the soot particles from each other. Conversely, it is likely that, at elevated temperatures, the average kinetic energy of the particles, $E = 3k_B(T + 273.15°)/2$, takes over the van der Waals adhesion energy, allowing the rearrangement of the mutual position of the particles. The thermophoretic force field might then bias these rearrangements so they have the effect of migrating the particles outward from the tube's surface. At the tips of the pyramids, the particles have a relatively small number of neighbors and can be "evaporated" into the hot gas surrounding the tube. This scenario is in line with current theories about the formation of nanoparticle aggregates, which assume that the coagulation efficiency of the particles depends on both their size and temperature.[38,39] At sufficiently high temperatures, smaller soot particles are likely to bounce off rather than adhere to each other because of the relatively high kinetic energy compared with the interaction energy of the particles (the so-called thermal rebound effect). Hou *et al.*[38] simulated the coagulation of soot particles in frames of the Hamaker model, in which the effective interaction of two spherical atomic clusters is the sum of all pairwise Lennard−Jones interactions of the atoms belonging to opposite clusters. As follows from their model, the van der Waals interaction cannot hold together two spherical soot particles of diameters $\leq 10$nm when the temperature exceeds $T \approx 2000°$C. These values are uncertain, since the soot particles are mostly aspherical at this temperature. Moreover, in close packing, each particle has more than one neighbor; therefore, it needs more kinetic energy to escape from them. Also, the formation of covalent bonds between some particles is not excluded. Nevertheless, it seems plausible that at high temperatures, which are reached in the resistively heated tubes, the soot particles may move with respect to each other, at least on the surface of the pyramids, where they have a smaller average number of contacts. It may, then, be possible to make an analogy between the ensemble of soot particles at high temperatures and a layer of suspended viscous fluid whose interface is deformed by gravity.[18] This analogy is supported by the existence of thin filaments by which the extremities of some pyramids are prolonged. Such filaments are unlikely to be formed by the random aggregation of soot particles. Rather, they look as the result of the viscous flow in a force field. Another argument for the liquid-like behavior of the soot layer on the surface of resistively heated microtubes is the circular growth of pyramid clusters (Figure S4). The already formed pyramids seem to induce the formation of subsequent generations of pyramids in their neighborhood. Similar behavior was observed for the arrays of pending droplets in the layers of suspended viscous fluid film,[18] in which the concentric rings of hexagonally coordinated droplet arrays grow around the initial perturbations.

Thermoelectric phenomena may be also relevant to the pyramids formation. A grounded carbon spherical nanoparticle of the diameter $a \approx 10$nm loses one electron every $\approx 110$ μs at $T = 2000°K$ ($\approx 1727°C$) and every $\approx 0.7$ μs at $T = 2400°K$ ($\approx 2127°C$), as calculated by the Richardson−Dushman equation for the thermoelectric current, $I \approx sAT^2 e^{-\Phi/k_B T}$, where $s = \pi a^2$ is the surface of the particle, $A = 1.2 \cdot 10^6 A \cdot m^{-2} K^{-2}$ is the Richardson's constant, and $\Phi = 7.63 \cdot 10^{-19} J$ is the work function for carbon.[40,41] Filippov et al. considered the possibility of an electrostatic breakdown (Coulomb explosion) of soot agglomerates heated by laser pulses to high temperatures.[40] If the analogy between the soot layer at high temperature and a layer of viscous liquid having a finite surface tension is correct, then charging particles on the surface of the layer may lead to the development of electrohydrodynamic instability.[42] However, unlike the particles in the isolated soot clusters, the particles on the surface of the tubes constantly restore their electroneutrality due to the electron flux from the grounded tubes. Unfortunately, in the absence of literature and data on the conductivity and effective permittivity of the soot layer at high temperatures, it is impossible to make reliable estimates of the surface charge.

Theoretically, it might be possible to discriminate, which of the two factor is more relevant to the pyramids formation, by heating the carbonized tubes in an oven, thus eliminating the temperature gradient at the tubes surface. But, the maximal temperature, at which the alumina ovens can operate in continuous regime does not exceed 1600° - 1700°C, which is essentially below of the lowest temperature, at which we succeeded to observe the pyramids. The graphitic ovens can reach much higher temperatures (up to 3000°C), but they are incompatible with the oxygen-containing atmosphere.

The arrays of the carbon micropyramids may be explored as the field emitter arrays.[9,10,43] As a first proof of concept we present here preliminary results to test the capacity of field emission properties of individual pyramids. We used a micro-nanoprobing station mounted in a scanning electron microscope, see Figure 6a. The nanoprobes are standard tungsten tips without any coating (such as for example gold to improve the workfunction of the tip probe). The measurements are taken at $10^{-6}$ mbar at the ambient temperature. An example of current-voltage (I-V) characteristic of a pyramid, which is separated from the tip by a 200nm wide gap, is shown in Figure 6b. The observed emission curves are very similar to those observed in a previous work devoted to the emission from carbon nanotube films and approximated by the classical Fowler–Nordheim (FN) model.[43] The first few sweeps show a noisy curves (blue, red, green lines).

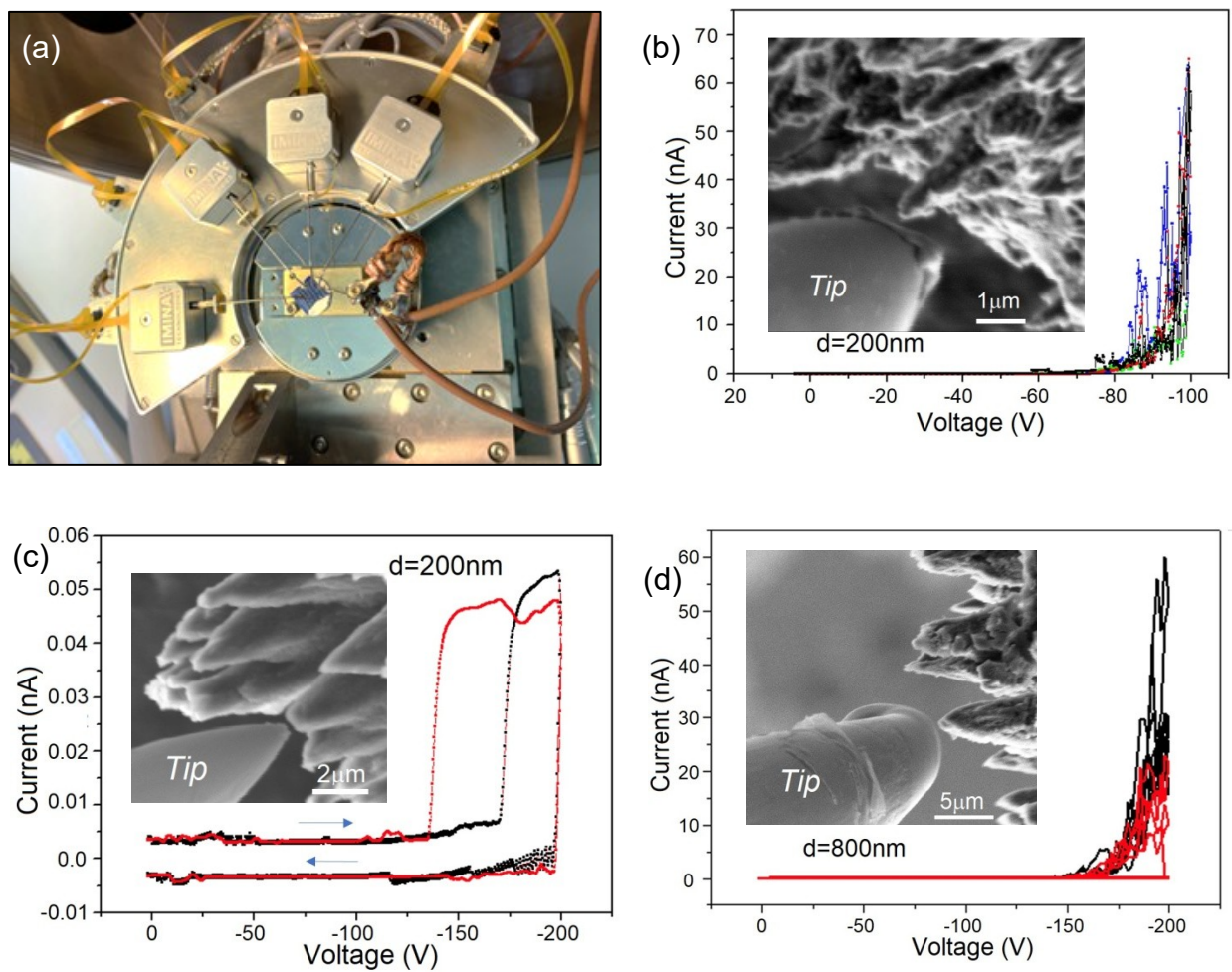

**Figure 6.** The measurement of the tunneling current from the micropyramids. (a) The set-up: four micro-nano-manipulator IMINA® mounted around a cryo stage (600K-77K), Kammrath® (b) SEM image of the nanoprobe in front of a pyramid (d=200nm) and the emitted current vs applied potential. The current increases with a threshold bias of -75V. (c) SEM image and anomalous (non-FN) emitted current versus bias curve for two pyramids. (d) The I-V curve for a pyramide annealed at 2300°C.

After several sweeps the I-V characteristics became regular (black lines). This is usually attributed to a "conditioning" of the emitter, such as oxygen desorption.[43] The threshold of electron emission is -75V and the measured emitted current reach 60nA for a bias of -100V. It should be said that the emissive characteristics of the pyramids are highly variable and may demonstrate completely different I-V curves. The Figure 6c shows two type of I-V curve recorded for two other pyramids. The shape of the curve is very different than the expected FN model. The current increases slowly up to a sharp increase at different bias voltage with a saturation at 0.5nA which is a value of one order of magnitude lower than for the pyramid shown on the Figure 6b. Possibly, the morphology of the pyramids changes during the I-V measurement, however a plateau of emissive current is reached. The FN-like emission curves were detected for the micropyramids annealed at 2300°C (Figure 6d).

From these measurements follows that the carbon micropyramids do can be considered as the field emission sources, yet their emissive characteristics are very variable. It is not surprising since, according to the FN theory, the tunneling currents are highly dependent on the shape of the extremities of the emitters (mathematically, it is expressed by the local field enhancement factor), and the extremities of the pyramids are all different. We need much more studies to understand the behavior of individual pyramids, to probe the emissivity with lower tip radius curvature (these experiments were done with 500nm but tip with 100 and 50 nm) to see the effect of the pyramids shapes and morphology, as well as the coating on the tungsten tip.

**Conclusion and outlook.**

Hexagonal arrays of micropyramids composed of soot nanoparticles are observed on the surface of graphitized microtubes resistively heated to $1800°C - 2400°C$ in the Ar atmosphere containing a small (~300-600p.p.m.) fraction of oxygen. The few micrometer tall and wide micropyramids are arranged in quasi-regular hexagonal arrays on the surface of the microtubes. The pyramids are pointing normally to the walls of the craters, except the borders between the craters, where the normal direction is not defined. The pyramids are formed only on the external walls of the microtubes. HRTEM reveals that the increasing fraction soot particles is morphologically identical to the carbon nano-onions, when resistive heating temperature grows.

The pyramids are soft, and can be easily destroyed by touching them, but they can be rigidified by the resistive heating without oxygen admixing in the Ar stream passing through the system. The mechanism of the pyramids formation and their self-assembly in the quasi-hexagonal arrays is not yet clear. The fact that the pyramids are always directing parallel to the local normals to the tube surface indicates on the existence of some factor which favors the drift of the nanoparticles in these directions. The thermophoretic force, arising as the consequence of a strong temperature gradient in the vicinity of the tube, is a natural candidate on the role of such a factor. Under this assumption, one can draw an analogy with the gravity-driven Rayleigh−Taylor instability of a thin suspended liquid layer. It is also possible that the electrostatic charging of the soot particle layer due to thermoelectric emission contributes to the pyramids formation and their assembly in the hexagonal arrays.

Regular mutual arrangement and the sharp extremities of the micropyramids allow to consider them as perspective micro-structures for advanced applications. In our study, we have demonstrated that the individual carbon micropyramids emit the tunneling current upon application of local electrical field. This indicates on the principal possibility to create field emitter arrays[45] on the base of the micropyramids ensembles. To achieve this goal, a reliable approach should be found for the formation of the pyramids arrays on the flat substrates and on sufficiently large surfaces. The pyramids ensembles on the flat substrates may be then investigated also for the creation of superhydrophobic[15] and bactericidal[16] coatings, as well as the ultra-black materials.[14]

**Methods.**

The experimental setup for the generation of the soot micropyramid arrays is shown schematically in Figure 7. An electro-conductive hollow-core microtube derived from a chitosan self-rolled microtube, made according to the procedure given in Ref. [44], was placed inside quartz vessel *1*, where it was suspended on crocodile-like electrodes. An electrical current was generated in the tube by a DC power supply *2*. The vessel was filled with an air/argon mixture at normal pressure $p = 101.3$ kPa with the use of valves *3* and *4*, guided by a multichannel gas flow controller *5*.

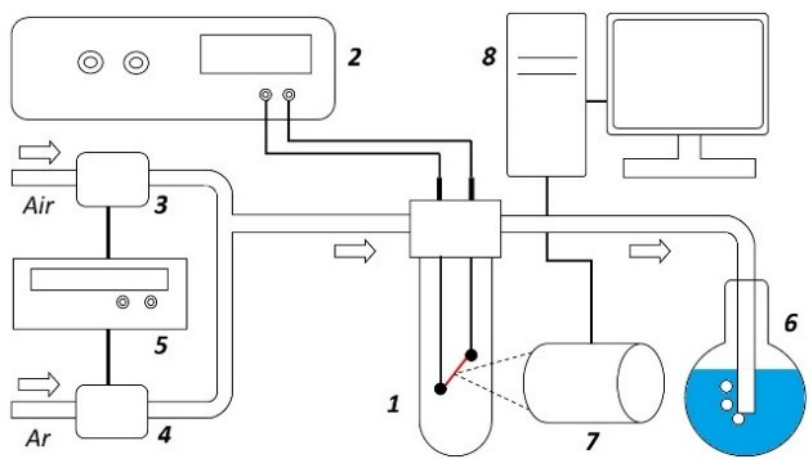

**Figure 7.** The experimental setup to produce arrays of self-assembled soot micropyramids.

After passing through the system, the gas mixture was rejected in atmosphere *6*. The relative concentration, $C_{O_2}$, of oxygen was varied between 300 and 3,000 p.p.m. The temperature of the microtube was measured by Modline 5 Raytec optical pyrometer *7* conjugated with a computer *8*. Typically, the formation of well-developed pyramid patterns requires several dozens of minutes $C_{O_2} = 300 p.p.m.$, but at a relatively high $O_2$ concentration, this time period was shorter. After a joule heating experiment, the filament was gently dismounted and examined by scanning electron microscopy (SEM) and transmission electron microscopy (TEM). Soot particles transferred onto Scotch tape were examined by Raman spectroscopy (Horiba Labram 300, λ=632nm). The field emission was measured with the use of a nanoprobing station mounted in a scanning electron microscope FEG-MEB XL30. The station consisted of four micro-nanomanipulators IMINA® mounted around a Kammrath® cryostage (600K-77K). Panasonic pyrolytic graphite sheets were purchased at Farnell Electronics.

**Acknowledgements**

We thank L. Vidal, S. Knopf, B. Rety, and A. Beda (all-IS2M CNRS) for their assistance with electron microscopy imaging and the carbonization of chitosan microtubes. The measurements of the tunneling currents were done with the financial support by the CNRS, the Région Grand Est and FEDER (EU) through the NanoteraHertz project, and the French National Research Agency (ANR) through the MIXES project (Grant ANR-19-CE09-0028).

**Supporting Information.**

The Supporting Information Available :

**Figure S1**, a control experiment on resistive heating of a tube without air admixing.

**Figure S2**, high resolution SEM images of the micropyramids.

**Figure S3**, a control experiment, etching of a pyrolytic graphite film.

**Figure S4**, the grow of "craters" on the surface of the resistively heated microtubes.

**Figure S5**, the morphology of a microtube, resistively heated at $T = 2000°C$ and oxygen relative concentration $C_{O_2} \approx 600$ p.p.m.

**Figure S6**, graphene flowers grown on the surface of a tungsten wire suspended in the vicinity of a resistively heated tube.

Manuscript accepted to ACS Nano
https://pubs.acs.org/doi/10.1021/acsnano.2c04395*Adv.* **2016**, *6*, 16745-16750.

[14] Sun, Y.; Evans, J.; Ding, F.; Liu, N.; Liu, W.; Zhang, Y.; He, S. Bendable, Ultra-Black Absorber Based on a Graphite Nanocone Nanowire Composite Structure. *Opt. Express.* **2015**, *23*, 20115-20123.

[15] Meng, L.-Y.; Park, S.-J. Superhydrophobic Carbon-Based Materials: a Review of Synthesis, Structure, and Applications. *Carbon Lett.* **2014**, *15*, 89-104.

[16] Fisher, L.; Yang, Y.; Yuen, M.-F.; Zhang, W.; Nobbs, A.; Su, B. Bactericidal Activity of Biomimetic Diamond Nanocone Surfaces. *Biointerfaces.* **2016**, *11*, p. 011014 (6pp).

[17] Cowley, M. D.; Rosenweig, R. E. The Interfacial Stability of a Ferromagnetic Fluid. *J. Fluid Mech.* **1967**, *30*, 671-688.

[18] Fermigier, M.; Limat, L.; Wesfried, J. E.; Boudinet, P.; Quiliet, C. Two-Dimensional Patterns in Rayleigh-Taylor Instability of a Thin Layer. *J. Fluid Mech.* **1992**, *236*, 349-382.

[19] Mora, S.; Phou, T.; Fromental, J.-M.; Pomeau, Y. Gravity Driven Instability in Elastic Solid Layers. *Phys. Rev. Lett.* **2014**, *113*, 178301 (5pp).

[20] Schäffer, E.; Thurn-Albrecht, T.; Russel, T. P.; Steiner, U. Electrically Induced Structure Formation and Pattern Transfer. *Nature.* **2000**, *403*, 874-877.

[21] Chakrabarti, A.; Mora, S.; Richard, F.; Phou, T.; Fromental, J.-M.; Pomeau, Y.; Audoly, B. Selection of Hexagonal Buckling Patterns by the Elastic Rayleigh-Taylor Instability. *J. Mech. Phys. Solids.* **2018**, *121*, 234-257.

[22] Waldmann, L. Uber die Kraft Eines Inhomogenen Gases auf Kleine Suspendierte Kugeln. *Z. Naturforsch.* **1959**, *14*, 589–599.

[23] Bacsa, W. S.; de Heer, W. A.; Ugarte, D.; Châtelain, A. Raman Spectroscopy of Closed-Shell Carbon Particles. *Chem. Phys. Lett.* **1993**, *211*, 346-352.

[24] Pujals, D. C.; de Fuentes, O. A.; Garcıa, L. F. D.; Cazzanelli, E.; Caputi, L. S. Raman Spectroscopy of Polyhedral Carbon Nano-Onions. *Appl. Phys. A.* **2015**, *121*, 1339-1345.

[25] Bockhorn, H. Soot Formation in Combustion. Springer: Berlin/Heidelberg, 1995.

[26] Frenklach, M. Reaction Mechanism of Soot Formation in Flames. *Phys. Chem. chem. Phys.* **2002**, *4*, 2028-2037.

[27] Hagen, F. P.; Rinkenburger, A.; Gunther, J.; Bockhorn, H.; Niessner, R.; Suntz, R.; Loukou, A.; Trimis, D.; Haisch, C. Spark-Discharge Generated Soot: Varying Nanostructure and Reactivity Against Oxidation with Molecular Oxygen by Synthetic Conditions. *J. Aerosol Sci.* **2020**, *143*, 43-
23

Manuscript accepted to ACS Nano
https://pubs.acs.org/doi/10.1021/acsnano.2c0439524Manuscript accepted to ACS Nano
https://pubs.acs.org/doi/10.1021/acsnano.2c0439558.

[28] Ajtai, T.; Utry, N.; Pintér, M.; Kiss-Albert, G.; Puskas, R.; Tapai, C.; Kecskeméti, G.; Smausz, T.; Hopp, B.; Bozoki, Z.; Konya, Z.; Szabo, G. Microphysical Properties of Carbonaceous Aerosol Particles Generated by Laser Ablation of a Graphite Target. *Atmos. Meas. Tech.* **2015**, *8*, 1207-1215.

[29] Hopf, C.; Angot, T.; Aréou, E.; Dürbeck, T.; Jacob, W.; Martin, C.; Pardanaud, C.; Roubin, P.; Schwarz-Selinger, T. Characterization of Temperature-Induced Changes in Amorphous Hydrogenated Carbon Thin Films. *Diam. Relat. Mater.* **2013**, *37*, 97-103.

[30] Abrahamson, J. Graphite Sublimation Temperatures, Carbon Arcs and Crystallite Erosion. *Carbon.* **1974**, *12*, 111-141.

[31] Kane, J. J.; Contescu, C. I.; Smith, R. E.; Strydom, G.; Windes, W. W. Understanding the Reaction of Nuclear Graphite with Molecular Oxygen: Kinetics, Transport, and Structural Evolution. *J. Nucl. Mater.* **2017**, *493*, 343-367.

[32] Ermakov, V. A.; Alaferov, A. V.; Vaz, A.; Perim, E.; Autrero, P. A.; Paupitz, R.; Galvao, D. S.; Moshkalov, S. A. Burning Graphene Layer-By-Layer. *Sci. Rep.* **2015**, *5*, 11546 (8pp).

[33] Bich, E.; Millat, J.; Vogel, E. The Viscosity and Thermal Conductivity of Pure Monatomic Gases from Their Normal Boiling Point up to 5000 K in the Limit of Zero Density and at 0.101325 MPa. *J. Phys. Chem. Ref. Data.* **1990**, *19*, p. 1289.

[34] Langmuir, I. Convection and Conduction of Heat in Gases. *Phys. Rev.* **1912**, *34*, 401-422.

[35] Devonshire, R.; Dring, I.; Hoey, G.; Porter, F. M.; Williams, D. R.; Greenhalgh, D. A. Accurate CARS Measurement and Fluid-Flow Modelling of the Temperature Distribution Around a Linear Incadescent Filament. *Chem. Phys. Lett.* **1986**, *129*, 191-196.

[36] Boutebba, S.; Kaabar, W. Simulation of Natural Convection Heat Transfer of Nitrogen in a Cylindrical Enclosure. *Period. Polytech. Mech. Eng.* **2016**, *60*, 256-261.

[37] Liu, Y.; Song, C.; Lv, G.; Chen, N.; Zhou, H.; Jing, X. Determination of the Attractive Force, Adhesive force, Adhesion Energy and Hamaker Constant of Soot Particles Generated From a Premixed Methane/Oxygen Flame by AFM. *Appl. Surf. Sci.* **2018**, *433*, 450-457.

[38] Hou, D.; Zong, D.; Lindberg, C.; Kraft, M.; You, X. On the Coagulation Efficiency of Carbonaceous Nanoparticles. *J. Aerosol Sci.* **2019**, *140*, 18-29.

[39] Wang, H. C.; Kasper, G. Filtration Efficiency of Nanometer-Size Aerosol Particles. *J. Aerosol Sci.* **1991**, *22*, 31-41.

[40] Filippov, A. V.; Markus, M. W.; Roth, P. In-Situ Characterization of Ultrafine Particles by Laser-

TOC figure

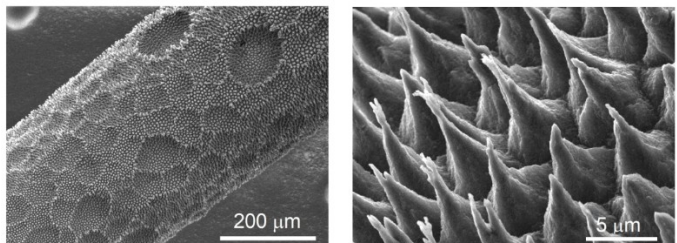

Self-assembly of soot nanoparticles into regular arrays of micropyramids was observed on the surface of carbon microtubes resistively heated in oxygen-deficient atmosphere.

**Self-assembly of soot nanoparticles on the surface of resistively heated carbon microtubes in near-hexagonal arrays of micropyramids -** *Supporting Information.*

*Valeriy A. Luchnikov,[1*] Yukie Saito,[2*] Luc Delmotte,[1] Joseph Dentzer,[1] Emmanuel Denys,[1] Vincent Malesys,[1] Ludovic Josien,[1] Laurent Simon,[1] Simon Gree.[1]*

[1]Université de Haute-Alsace, CNRS, IS2M, UMR 7361, Mulhouse F-68057, France
[2]Graduate School of Agricultural and Life Sciences, The University of Tokyo, Tokyo, 113-8657, Japan

*Corresponding authors: Valeriy A. Luchnikov, email: valeriy.luchnikov@uha.fr and Yukie Saito, email: aysaito@g.ecc.u-tokyo.ac.jp

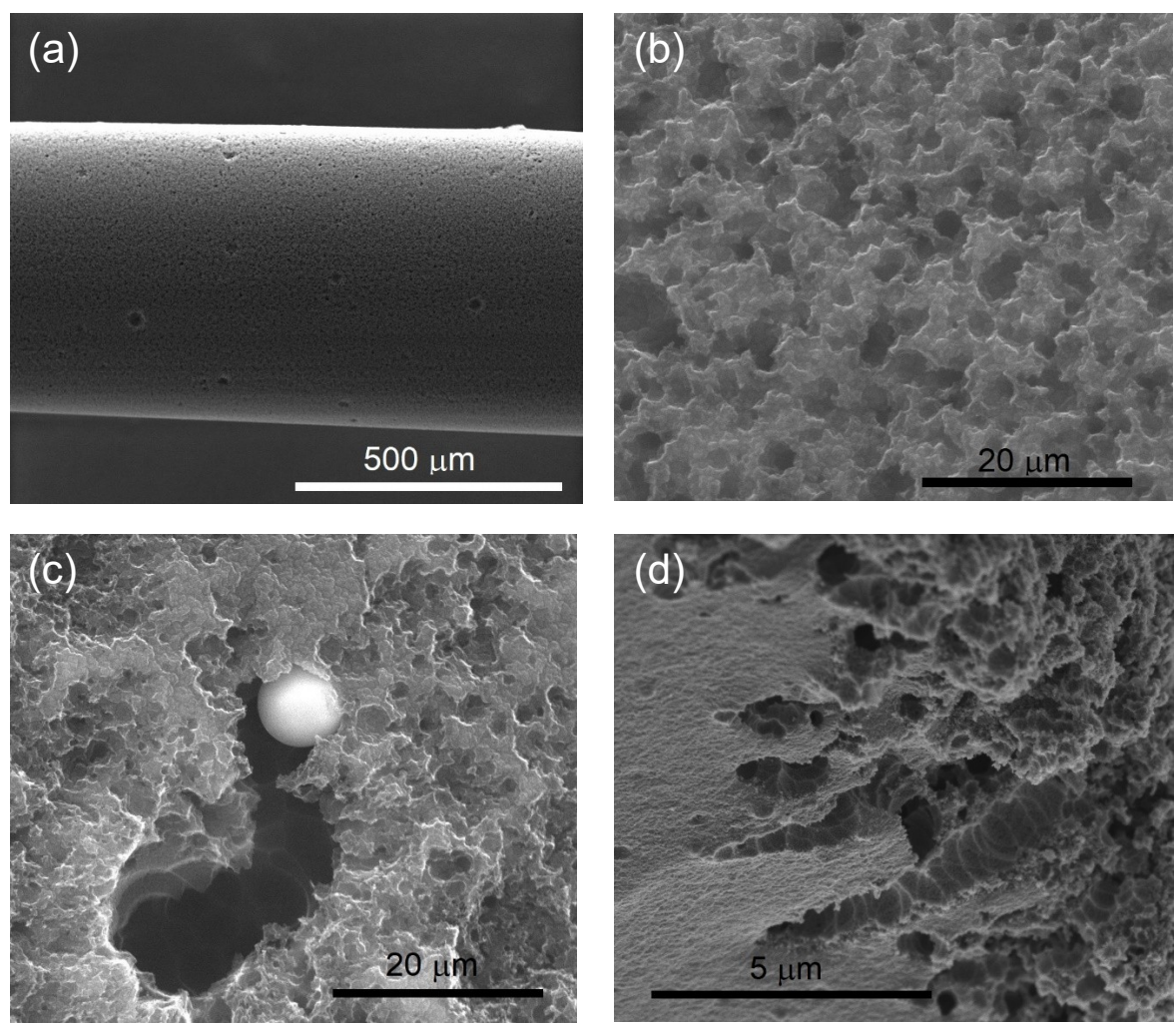

**Figure S1.** Control experiment on resistive heating of a tube without air admixing, at $T = 2000°C$. (a,b) Surface of the tube, at different magnifications. (c) An impurity escaping the microtube surface. (d) Cross-section view of a surface of a joule-heated microtube, showing the traces of impurities which have migrated on the surface and which are responsible for the surface roughness.

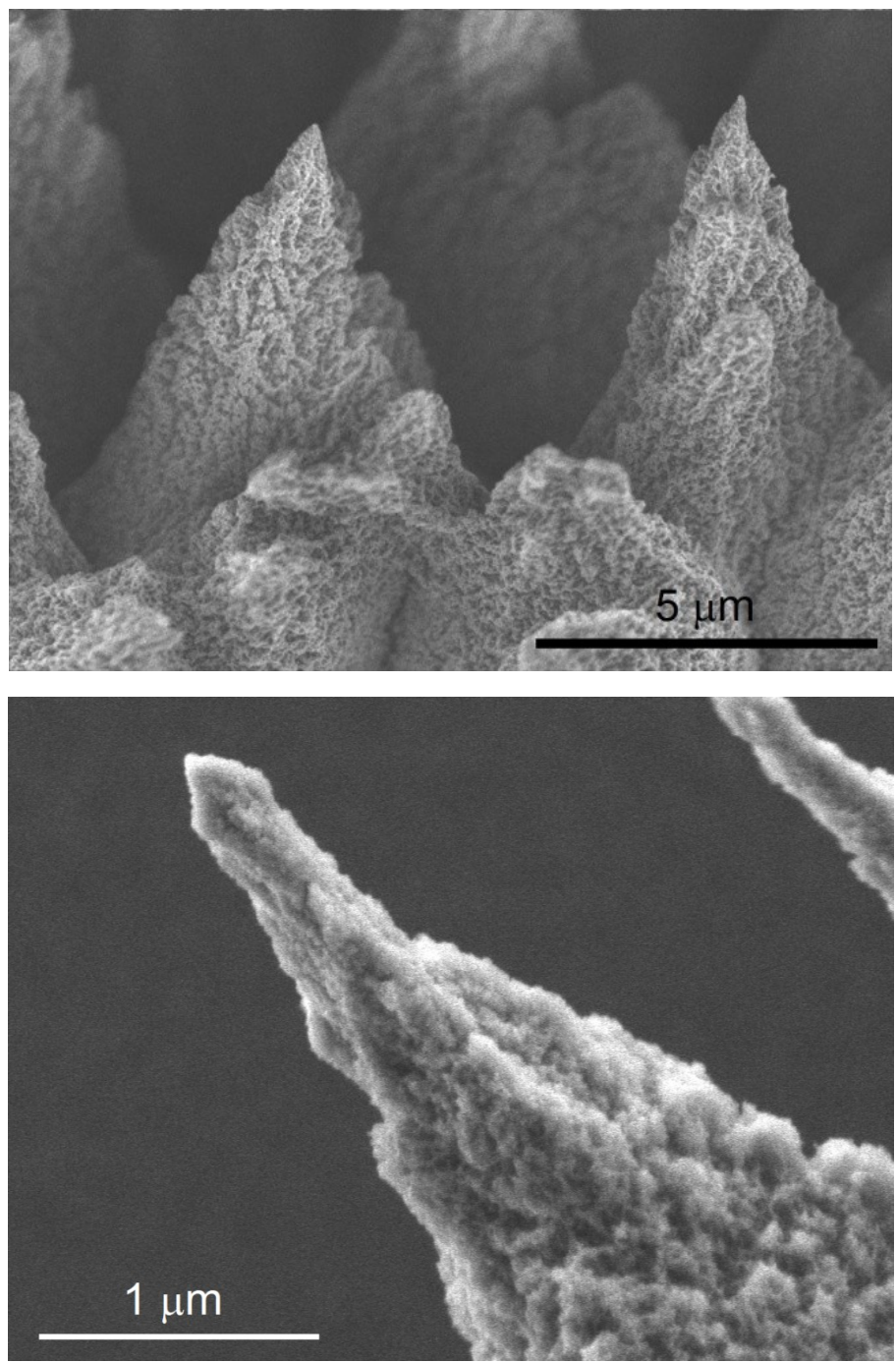

**Figure S2.** Micropyramids, obtained at $T = 2000°C$ (high resolution SEM).

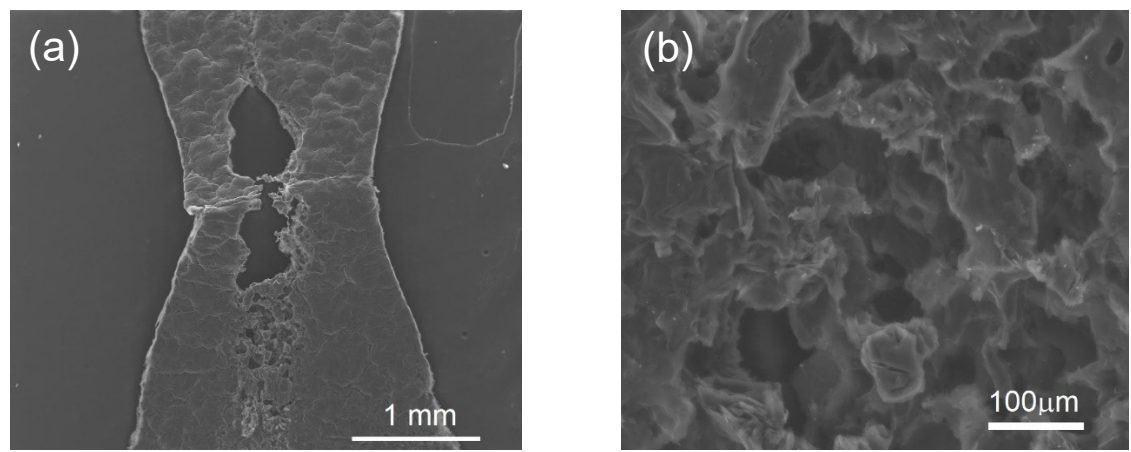

**Figure S3.** The control experiment – etching of a pyrolytic graphite film at 2000°C and oxygen relative concentration 600p.p.m. during 20 min. No soot formation is detected.

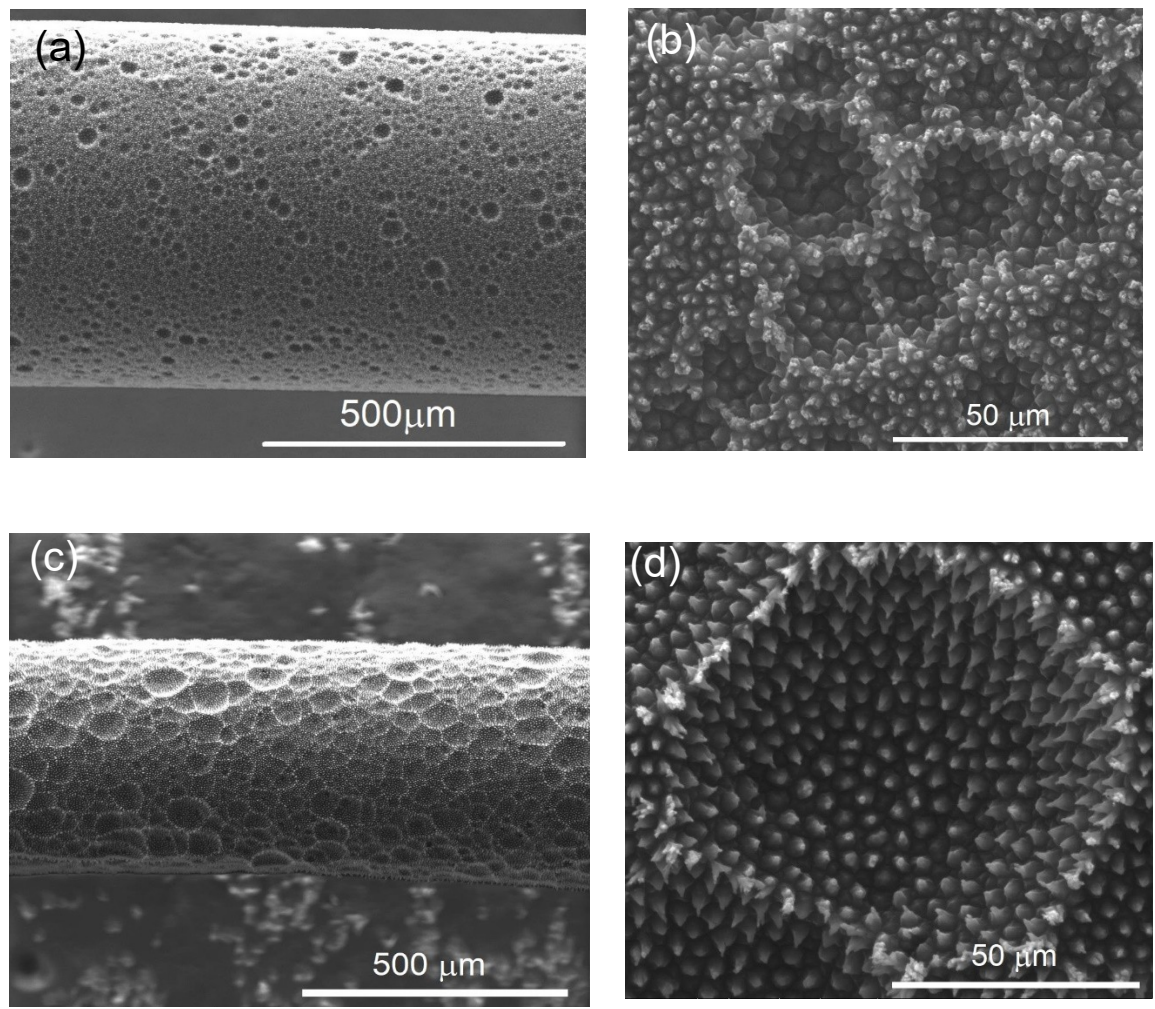

**Figure S4.** Grow of the "craters" on the surface of the microtubes resistively heated at $T = 2000°C$ and oxygen relative concentration $C_{O_2} \approx 300\ p.p.m.$ (a,b): a tube surface after 80 min of heating. (c,d) the same

tube surface, after 320 minutes of heating.

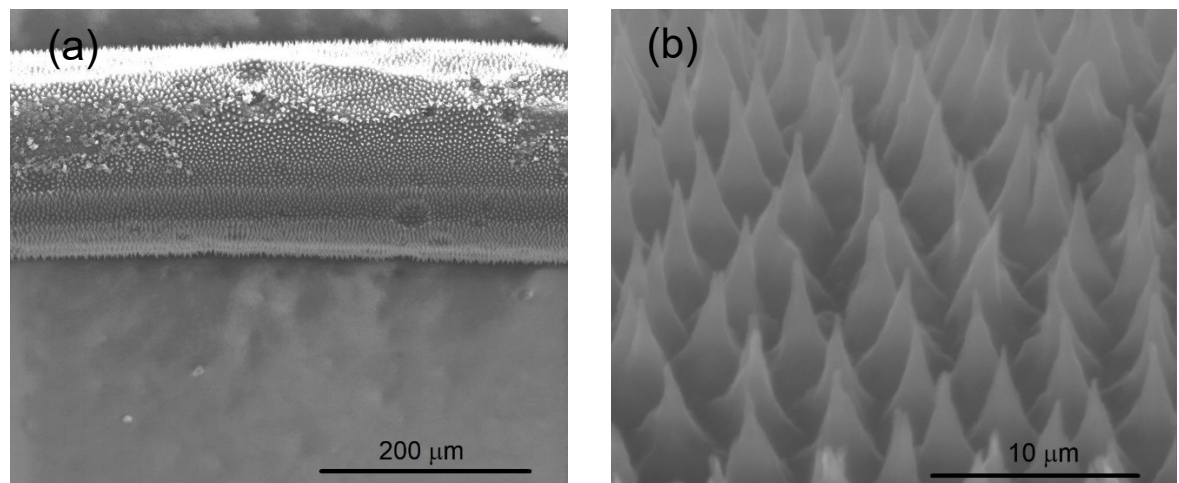

**Figure S5.** The morphology of a microtube, resistively heated at $T = 2000°C$ and oxygen relative concentration $C_{O_2} \approx 600\ p.p.m.$ (after 120 minutes of heating).

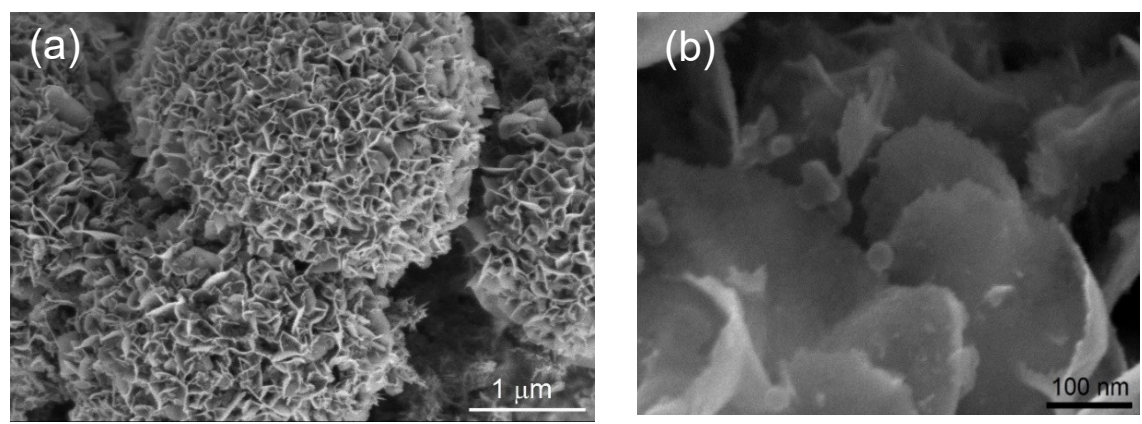

**Figure S6.** (a) Graphene flowers grown on the surface of a tungsten wire suspended in a carbon tube resistively heated to 2300°C. (b) Soot particles between the "petals" of the GF.